\documentclass[preprint,showpacs,preprintnumbers,amsmath,amssymb,
               pra,floatfix]{revtex4}

\usepackage{graphics}% Include figure files
\usepackage{bm}% bold math

\begin{document}

\title{ Resonance and virtual bound state solutions of the radial
        Schr\"odinger equation }

\author{I.\ Borb\'ely}
\affiliation{Research Institute for Particle and Nuclear Physics,
         H-1525 Budapest P.O. Box 49,
         Hungary }

\date{\today}

\begin{abstract}
The formulation of the eigenvalue problem for the Schr\"odinger equation is 
studied, for the numerical solution a new approach is applied. With the 
usual exponentially rising free-state asymptotical behavior, and also with a 
first order correction to it, the lower half of the $k$ plane is
systematically explored. Note that no other method, including the complex 
rotation one, is suitable for calculating virtual bound states far from the 
origin. Various phenomena are studied about how the bound and the virtual 
bound states are organized into a system. Even for short range potentials, 
the free state asymptotical condition proves to be inadequate at some 
distance from the real $k$ axis; the first order correction doesn't solve 
this problem.

Therefore, the structure of the space spanned by the mathematically
possible solutions is studied and the notion of minimal solutions is 
introduced. Such solutions provide the natural boundary condition for 
bound states, their analytical continuation to the lower half plane
does the same for resonances and virtual bound states. The eigenvalues
coincide with the poles of the S-matrix. The possibility of
continuation based on numerical data is also demonstrated. The
proposed scheme is suitable for the definition of the in- and outgoing
solutions for positive energies; therefore, also for the calculation
of the so called distorted S-matrix in case of long range
interactions.\end{abstract}

\vspace*{1cm}

\maketitle

\section{\label{sint} Introduction}

Description of resonances in scattering processes is an inherent part
of quantum mechanics \cite{N66}, many publications are studying them.
The natural way to define a resonance or virtual bound state is
provided by the poles of the scattering amplitude in the complex
energy variable. Usually, for numerical reasons it is possible to
calculate resonances only with a small imaginary part of the energy
i.e. the so called "narrow" resonances. To overcome this limitation
the complex rotation method is used\cite{Sim73,Mois98}. With a
non-uniter transformation the energy spectrum is deformed in such a
way that any given resonance becomes part of the point spectrum and
with proper numerical method its energy can be determined. But the
applicability of the method is limited, it is not possible to
calculate virtual bound states. In this respect c.f. the explicit
statement made in ref.\cite{Ser01}: "Because complex rotation methods
cannot give the virtual states, we use a numerical integration of the
Schr\"odinger equation".

In the present paper the formulation of the eigenvalue problem 
directly for the Schr\"odinger equation is studied. For resonances and
virtual bound states not only the numerical method to solve the
equation is unstable, but also the applied boundary condition at
infinity proves to be problematic.

Bound states are defined by two requirements: the wavefunction is
regular at the origin and exponentially decreases at infinity. These
independent conditions are satisfied only at specific energies. The
standard method for finding the eigenvalues can be called "midpoint
matching". The Schroedinger equation is integrated in two steps: in
the internal region from the origin up to a suitably chosen matching
point $R_m$ and in the external region from a large radius $R_M$
inward to $R_m$. Since the unwanted other solutions decrease in the
chosen directions, the numerical results are accurate and one can
easily find those energies where the two solutions concurrent with
each other. Naturally, the existence of the $R_M \rightarrow \infty$
limit has to be examined.

Unfortunately, for resonances and for virtual bound states the method
does not work in this form. The integration in the internal region
remains the same, but since these eigenvalues are defined by an
exponentially \emph{increasing} asymptotical behavior, the inward
integration in the external region is inaccurate: any small admixture
of the asymptoticly decreasing solution catastrophically increases.
The loss of accuracy is not critical only if the imaginary part of the
wave-number $k$ is small, in other cases  one has to restrict the
calculation to very limited values of $R_M$.

A new effective approach to overcome
the above limitation is proposed in the present paper.
In its first form this method was proposed in
ref.\cite{BV95}, but here it is developed further.
As an example, it is applied with a short range potential, the
Saxon-Woods one, but general problems are addressed. Though the single
channel case is considered to be completely understood, the reliable
results provided by the numerical method make it necessary to solve
some principal problems. Namely, it is necessary to clarify the proper
asymptotical behavior of the wavefunctions. The usually supposed
"exponentially rising free-state solution" is only an approximation,
valid only for short range potentials near the real $k$ axis.

The cornerstone of the constructions presented is the
notion of minimal solutions, a similar notion is used in the theory of
three-term recurrence relations \cite{JT80}. Since it provides a
concise formal way to describe the exponentially decreasing solutions,
it is introduced on a relatively early stage. In the two-dimensional
linear space spanned by the mathematically possible solutions at a
fixed energy, the minimal solutions are those solutions  which
decrease faster at
infinity than the other ones.  The minimal solutions form a
one-dimensional subspace at energies outside of the positive axis
(on the half-axis they obviously do not exist), they provide the
natural asymptotical condition for the possible bound states and
because of their behavior they can be easily computed.

One can overcome the difficulties present for the inward integration
for resonances and virtual bound states by changing the representation
of the wavefunction. If sufficiently accurate information on the
asymptotical behavior is available, then one can calculate
only the correction to this behavior. "Sufficiently accurate"
means that the
remainder term is smaller than a minimal solution. Such a
representation results in an inhomogeneous differential equation,
the behavior of the source term assures that the inward
integration is accurate \cite{BV95}. 
Since the asymptotical behavior is used as a reference function for the
solution, I propose to call the approach  "reference method".
In the present paper it is pointed out that the
formulated method provides accurate results for a wide class of
solutions rather than only for that single solution for which the
supposed asymptotical behavior is valid.

As usual, accurate numerical results reveal inaccuracies in
theoretical understanding: it is necessary to clarify the proper
boundary condition for resonances and virtual bound states. At first
the usual free-state asymptotical behavior was assumed. Using it 
an infinite row of resonances with a nearly constant imaginary part is 
detected. 
Their position depends on the range parameter
of the potential and on the distance where the free-state behavior is
imposed, therefore such resonances are clearly unphysical. Since the
free-state
boundary condition strictly corresponds to the truncation of the
potential, one can call them truncation eigenvalues. To "improve" the
boundary condition, a first order correction generated by the exponential 
asymptotics of the potential is calculated. But the unphysical
resonances survived with being pushed downwards. Since the numerical
method itself is accurate, the usually neglected subtle problem of the
correct boundary condition has to be solved in a general way.

But before that, with the above boundary conditions
various phenomena are studied. These include 
the dependence of the eigenvalue positions
on the potential parameters (including collisions), eigenvalues
corresponding to the deep potential well behind the centrifugal
barrier and depending practically only on the radius of the potential,
the node number rules for the virtual bound states, the
interrelation between the bound and virtual bound states and so on. 
The behavior of the eigenvalues in the range singularity generated 
by the exponential tail of the potential is also studied.
Despite the "approximative" nature of the used boundary conditions, 
important factual results are received.

It is by no means straightforward to
determine the asymptotical behavior for resonances and virtual bound
states in a general way. By using only the solution space at fixed
energy, it is not possible to find in a natural way somehow an "other"
solution which is different from the minimal one. Time reversal does
not help
and the only available mathematical structure, namely the symplectic
form provided by the Wronski determinant for the solutions, is not
sufficient to define an orthogonal subspace to the subspace spanned by
the minimal solutions. Analyzing the above presented methods to choose
the boundary condition, one can realize that their essence is the
continuation of some approximation, 
which is valid on the upper half-plane, to
the lower one. Note that the evaluation of a formula automatically
performs an analytical continuation. Therefore the proper condition is
provided by the  continuation of the "exact", i.e. the minimal,
solutions to the lower half of the complex wave-number plane. 
If the wavefunction at $R_m$ is continued, one can spare even the
inward integration.

In this context the analytical dependence of the solutions was
discussed and the feasibility of the continuation to the lower
half-plane was successfully demonstrated using data provided by the
reference method. Since the solutions are defined by both the value of
the wavefunction and its derivative, while the normalization is
irrelevant, the continuation is performed for a function with values
in what is called "projective complex plane".

The principle of continuation for the minimal solutions has an
important and unexpected consequence. Minimal solutions can be
calculated for any potential, including long range ones, without
detailed information on the asymptotics and their continuation to the
real $k$ axis provides the ingoing and outgoing solutions. It is
straightforward to describe the "physical solution", i.e. the solution
regular at the origin, as a linear combination of these solutions. The
coefficients determine what can be called the "distorted S-matrix". In
this way one can perform scattering calculations practically for any
long range potential, the only prerequisite condition is the existence of
minimal solutions outside the positive energy axis.

Finally, it is pointed out that, because of the connection between the
minimal and the in- and outgoing solutions, the eigenvalues
defined with the continued minimal solution boundary condition
coincide with the poles of the S-matrix.

Due to the complex nature of the studied problem, the present paper
contains quite a heterogeneous material: from numerics to complex
projective planes. Mostly only simple logical steps are used, but to
understand them the reader has to be acquainted with the corresponding
field. Numerical illustrations are often given, but in fact they play a
more important role, they serve as source of information on the
studied problem. In this respect numerics plays exactly the same role
as measurements do in an experimental study. As far as it was possible, 
abstract mathematical notions are not used in the interpretation,
since it is usual in nuclear physics.
But general notions are to express the very essence of the phenomena, 
therefore they are unavoidable for a deeper understanding.

The present paper is not about the history of the subject. It was not
traced back where the results appeared first, some
of them could have been received for exactly solvable cases.
On the whole, the approach is quite different
from the usual one, it is irrelevant if the priority for some
particular details belongs to others. 
The proposed numerical approach (i.e. the
reference method used in the first part of the paper), many of the
factual results concerning virtual bound states, the formulation of
the eigenvalue problem with the boundary condition provided by the
analytical continuation of the minimal solutions and the
implementation of this continuation for a function with values in
$\mathrm{P}\mathbb{C}$ are definitely new.

\section{\label{sprel} Preliminaries, some conventions used in the
paper}

The Schr\"odinger equation the solutions of which are sought is
\begin{equation}
 \varphi''(r)+\biggl( k^2-V(r)-\frac{L(L+1)}{r^2} \biggr)\varphi(r)=0
\end{equation}
with standard notation. For $V(r)$ the Saxon-Woods potential
is chosen as a typical example
\begin{equation}
 V(r)=\frac{V_o}{1+\exp((r-R_o)/a)}
\end{equation}

Such a form is used to describe the thickness of the
nuclear surface layer around $r=R_o$. But, as a side effect, the
potential has an exponential tail $W_o=V_o\exp(-\mu r)$, where
$\mu=1/a$ and $W_o=V_o\exp(R_o/a)$. Therefore the same $a$ parameter
determines the asymptotical behavior too. The phenomena studied in
this paper are mostly connected to the asymptotics.

If not stated otherwise, the following values are used for the
parameters: $R_o=7.5 \, \mathrm{fm}$, $\mu=1.4 \, \mathrm{fm}^{-1}$.
The value for $V_o$ is usually $-2.0 \,\mathrm{fm}^{-2}$, such a
potential is roughly realistic for a nucleon in the
${}^{208}\mathrm{Pb}$ nucleus.

As it is described in the introduction, for the calculation of the
eigenvalues a matching procedure is performed, for it the mostly used
parameter values are $R_m=R_o$ and $R_M=30.0 \, \mathrm{fm}$.

The differential equation uniquely defines the solution if at some,
i.e. necessarily finite, point the value of the solution and of its
derivative are given. But for the studied problem the boundary condition
is imposed at infinity. It is said that a given expression describes
the asymptotical behavior of the solution if with this boundary
condition imposed at $R_M$ the $R_M \rightarrow \infty$ limit exists
for the solution. With other words, it is implicitly supposed that the
remainder term is smaller than any  other
solution. The basic principal problem addressed in this paper is the
the correct asymptotics. For bound states there is no problem since 
the other solution decreases inwards, but for resonances and
virtual bound states it is preferable to formalize the description.

\section{\label{sstruct} Structure of the solution space, minimal
solutions}

At a given energy the solutions of the Schr\"odinger equation
are uniquely defined by the value of the wavefunction and its
derivative at a fixed point, the mathematically possible solutions
span a two-dimensional vector space. Provided they exists, the so
called minimal solutions\cite{JT80} define a one-dimensional subspace
in a natural way. An $m(r)$ solution is called minimal, if another
$d(r)$ dominant solution exists for which $m(r)/d(r)\rightarrow0$ at
$r\rightarrow\infty$. Any constant multiple of $m(r)$ is minimal,
any solution outside of the linear subspace defined by $m(r)$ is a
dominant pair to it. Moreover, if there are two solutions which are
minimal, then they are necessarily constant multiple of each other.
The proof is straightforward: independent $m_1(r)$ and $m_2(r)$
serve as a basis to expand the corresponding $d_1(r)$ and $d_2(r)$
functions, but this assumption gives contradicting conditions on
$m_1(r)/m_2(r)$.

It is well known that except for positive real energies minimal
solutions do exist. In this paper they are labeled by the wave-number
$k$ defined by the energy parameter, $E=k^2$. By definition, that sign
of the square root is chosen for which $\mathrm{Im}(k) > 0$.

Minimal solutions are of great theoretical and practical importance.
The usual condition imposed on the asymptotical behavior of a bound
state wavefunction assures that it is a minimal solution. Note also,
that the regularity condition in the origin also can be interpreted in
terms of minimal solutions (of an other type, of course). But what
makes minimal solutions really important is the unique property that
their value at a given point can be easily calculated by integrating
inward from a large distance with a nearly arbitrary boundary
condition: despite the presence of a contaminating other solution only
the minimal solution survives. While from a theoretical point of view
a minimal solution is uniquely defined, for the numerical practice it
can be thought of as a bundle of different solutions which all agree
to a certain accuracy in a given interval, but which can widely differ
at very large distances. In other words, for numerics the minimal
solution is a family of different exact solutions.

Resonances and virtual bound states have an asymptotical behavior
obviously different from the minimal solution. But it is not
straightforward to determine their behavior. Note that to speak of
the "exponentially rising  solution" is highly inaccurate, with any
admixture of the  minimal solution the same property holds. It is
always necessary to use some, perhaps implicit, assumption to
eliminate this ambiguity. Unfortunately it is not possible to find in
a natural way somehow an "other" solution, different from the minimal
one, using only the solution space at fixed energy. The principal
reason is that the only mathematical structure, namely the symplectic
form provided by the Wronski determinant for the solutions, is clearly
not sufficient to define an orthogonal subspace to the subspace
spanned by the minimal solutions. In fact, it is possible only in the
opposite direction: after one chooses a subspace, is it possible to
define a scaler product, like for Kaehler manifolds \cite{Simpl}.
Therefore at first the intuitively obvious, i.e. "the
exponentially rising" free-state solution is chosen 
and the symptoms provided by it are carefully analyzed. 
It is not a waste of efforts, during it
interesting results are obtained. Only after some information is
gathered is it possible to point out that the analytical continuation
of the minimal solutions in the energy variable is the principal
answer.

\section{\label{srefm} The reference method}

In this section an approach to solve the
radial equation in the external region is introduced. 
The general considerations
presented here are applied and therefore illustrated in the next
sections.

The basic idea of the method is quite simple: the solution of the
Schr\"oedinger equation is formally split into two terms
$\varphi(r)=f(r)+u(r)$, where $f(r)$ is a fixed reference function
and $u(r)$ is the correction function to be calculated. It yields in a
trivial way
\begin{equation}
u''(r)+\bigl( k^2-V(r)-\frac{L(L+1)}{r^2} \bigr)u(r)=s(r),
\end{equation}
with $s(r) \equiv -f''(r)-\bigl( k^2-V(r)-\frac{L(L+1)}{r^2}
\bigr)u(r)$.
By choosing $f(r)$ and the boundary condition for $u(R_M)$ in
various ways one can influence the behavior, and thus the possible
numerical accuracy of the $u(r)$ solution.

At the beginning it is supposed that for the $\varphi(r)$ solution 
the $\varphi_a(r)$ asymptotical behavior is known
with an accuracy better than any minimal solution. It means that the
$\bigl ( \varphi(r)-\varphi_a(r) \bigr )/m(r) \rightarrow 0$ relation
holds at $r \rightarrow \infty$ for some minimal solution $m(r)$.
Obviously, this condition is in agreement with the definition of
the asymptotical behavior made in section \ref{sprel}. In this case
the above scheme with $f(r)=\varphi_a(r)$ and with the
$u(R_M)=0, u'(R_M)=0$ (i.e with the
$\varphi(R_M)=\varphi_a(R_M), \varphi'(R_M)=\varphi'_a(R_M)$)
boundary condition can be applied.
Since $u(r)=\varphi(r)-\varphi_a(r)$, during an
inward integration the correction function rises faster than the
minimal solutions.

To discuss the propagation of numerical errors, recall that
the general solution of an inhomogeneous equation is a
particular solution (e.g. the solution with the $u(R_M)=0,u'(R_M)=0$
boundary condition) plus the general solution of the homogeneous
equation. One can think of a single numerical integration step as a
stochastic source of small solutions of the homogeneous equation,
while the further steps are accurate to the contaminated solution.
"Small" means that both the function and its derivative are small.
The introduced contamination can be represented as linear
combination of the complete solution $\varphi(r)$ and of $m(r)$.
It means that any possible error can propagate inwards at most as a
minimal solution. The presence of the $\varphi(r)$ admixture is not
dangerous, it slightly influences only the norm of the result. But in
the generic case even a small admixture of  $m(r)$ can be
catastrophic.

In the given case $u(r)$ is rising faster than the minimal solutions,
therefore the relative contribution of the contaminating solution is
fading out in case it is sufficiently small and it is considered
sufficiently far from the place of its admixture. Moreover, 
a similar statement
is valid for the relative error introduced by a
slightly inaccurate boundary condition for $u(r)$.
Therefore, in any point far from $R_M$, let us say at $R_m$,
the numerical error is defined only by the accuracy of the
nearby integration steps. With other words, in case of
$R_M \rightarrow \infty$ the limit for the solution exist.

The reference method is not a numerical one in the strict sense. Only
the differential equation is reformulated in order to influence
the behavior of that part of the solution which is numerically
calculated. The resulting inhomogeneous equation
can be solved with any available numerical method,in the present paper
the standard Runge-Kutta-Fehlenberg one was used.
When high accuracy was needed, a high order routine was chosen.

A great advantage of the proposed integration scheme is that
for any given reference function one can easily check in a empirical
way if there is a corresponding solution simply by comparing the
correction function to a minimal solution. With other words, there is
no need to derive the reference function in an exact way,
an \emph{a posteriori} check is possible.

One can also exploit the freedom provided by the boundary condition
for $u(R_M)$. Since the integration of the minimal solution
is stable, one can describe its presence by including it into the
boundary condition. If its relative weight is sufficiently large, than
at $R_m$ it is still present in the result (with a small numerical
error, of course). To get the $R_M \rightarrow \infty$ limit, however,
one has to choose its weight correspondingly. It means that it is
not necessary to adhere to the solution the asymptotical behavior of
which is defined by the used reference function, the space of
solutions for which the integration is stable is larger.
Note also that the minimal solution
component in the boundary condition can be replaced by any of the
numerically equivalent solutions mentioned in section \ref{sstruct}.

One can go even further and incorporate not just the minimal solutions
into the boundary condition. In this case the simplest approach is
to express the boundary condition in terms of the reference function
and a fixed minimal solution, consequently apply a
renormalization of the reference
function and therefore the source term too.  It makes possible to
accurately perform the inward integration for an arbitrary boundary
condition.

\section{\label{sfreeb} A naive approach: free-state boundary
condition}

At first the usual naive approach is adopted: it is supposed that the
asymptotical behavior of the virtual bound and resonant states is
provided by the "exponentially rising" solution of the free equation.
Namely, for $L=0$ the asymptotical behavior is given by $\varphi_a(r)
= \exp( i k r )$. In case of a short range potential and near the
real $k$ axis it is an adequate condition.
For the correction function
the reference method yields the inhomogeneous equation
\begin{equation}
u''(r)+(k^2-V(r)) u(r) = \exp( i k r ) V(r)
\end{equation}
It is assumed that the boundary condition is $u(R_M)=0, u'(R_M)=0$ at some
large but necessarily finite $R_M$.

\begin{figure}
\includegraphics{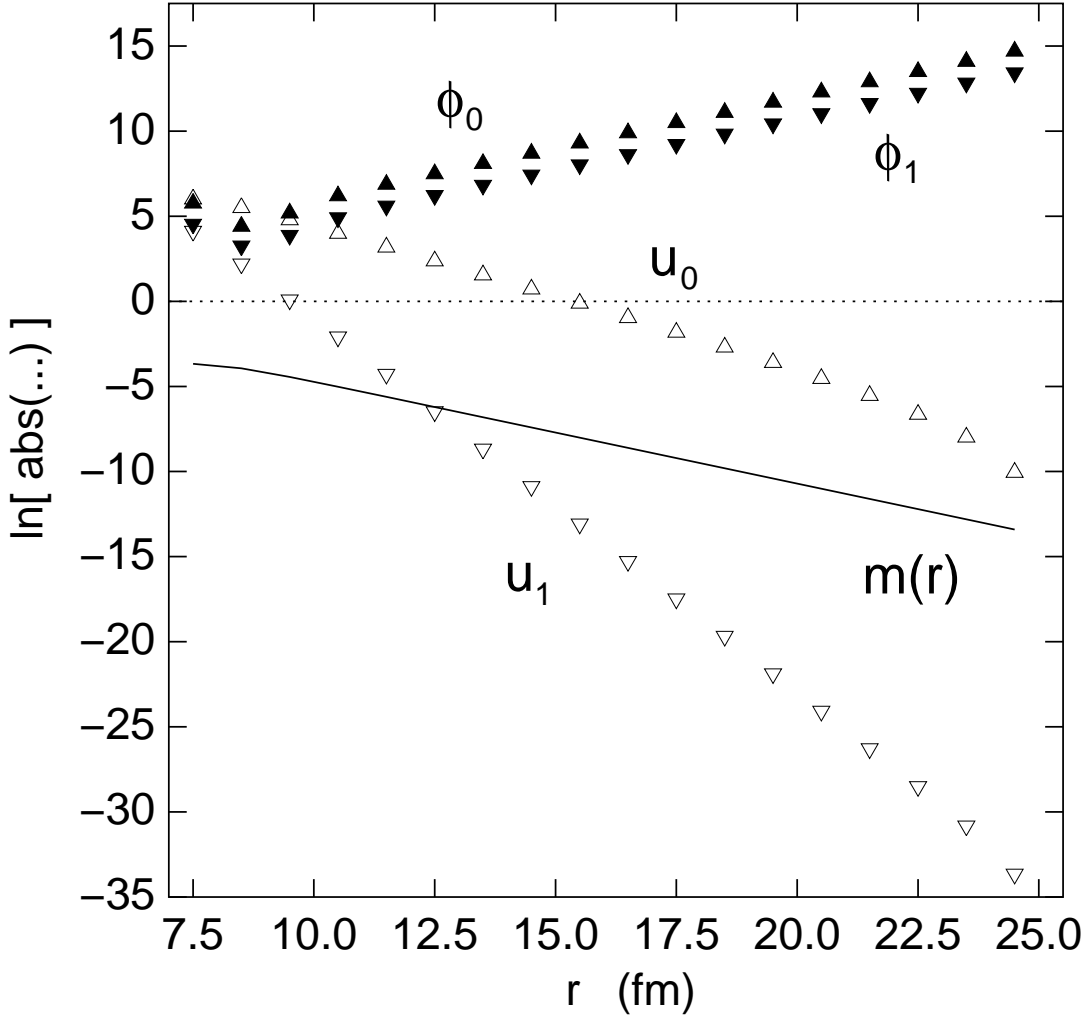}
\caption
{\label{fig1graph}
Stability of the integration schemes at $k=-i\cdot 0.6 \;
\mathrm{fm}^{-1}$. Free-state boundary condition results are given by
triangles, the corresponding functions are marked by a zero index, the
iterated boundary condition results are given by inverse triangles and
marked by index one. The open symbols give the correction functions,
the full ones the complete wavefunctions. The solid line gives a
minimal solution, any error can increase during its
propagation at most according to it. Due to the different boundary
conditions the complete solutions are different, but the rough
logarithmic scale hides it. Apart from the normalization, the
deviation between their values is the order of 4\%. }
\end{figure}

Details of the numerical integration for the external region at
$k=-i \cdot 0.6 \; \mathrm{fm}^{-1}$ are presented in
fig.\ref{fig1graph}. It can be
seen that the method is numerically stable: any error introduced at
some $r$ can propagate inwards rising at most as a minimal solution
$m(r)$. But due to the presence of the source term, $u(r)$ rises
faster. This proves \emph{a posteriori} that there is a solution of
the Scr\"oedinger equation which has the supposed asymptotics.

By taking into account that the asymptotical behavior of the solutions
is $\exp(\pm i k r)$, simple straightforward considerations as well as
an empirical check show that the stability region for the above
integration scheme is $\mathrm{Im}(k)>-\mu/2$. For $k$ values inside
this region the limit for $u(R_m)$ exists at $R_M \rightarrow \infty$,
therefore its value can be calculated to an accuracy defined by the
nearby integration steps only.

Note that the generalization of the above simple formulation to
arbitrary orbital momenta even in the presence of a Coulomb
interaction is straightforward: the known solutions for these cases
can be chosen as the reference function.

\begin{figure}
\includegraphics{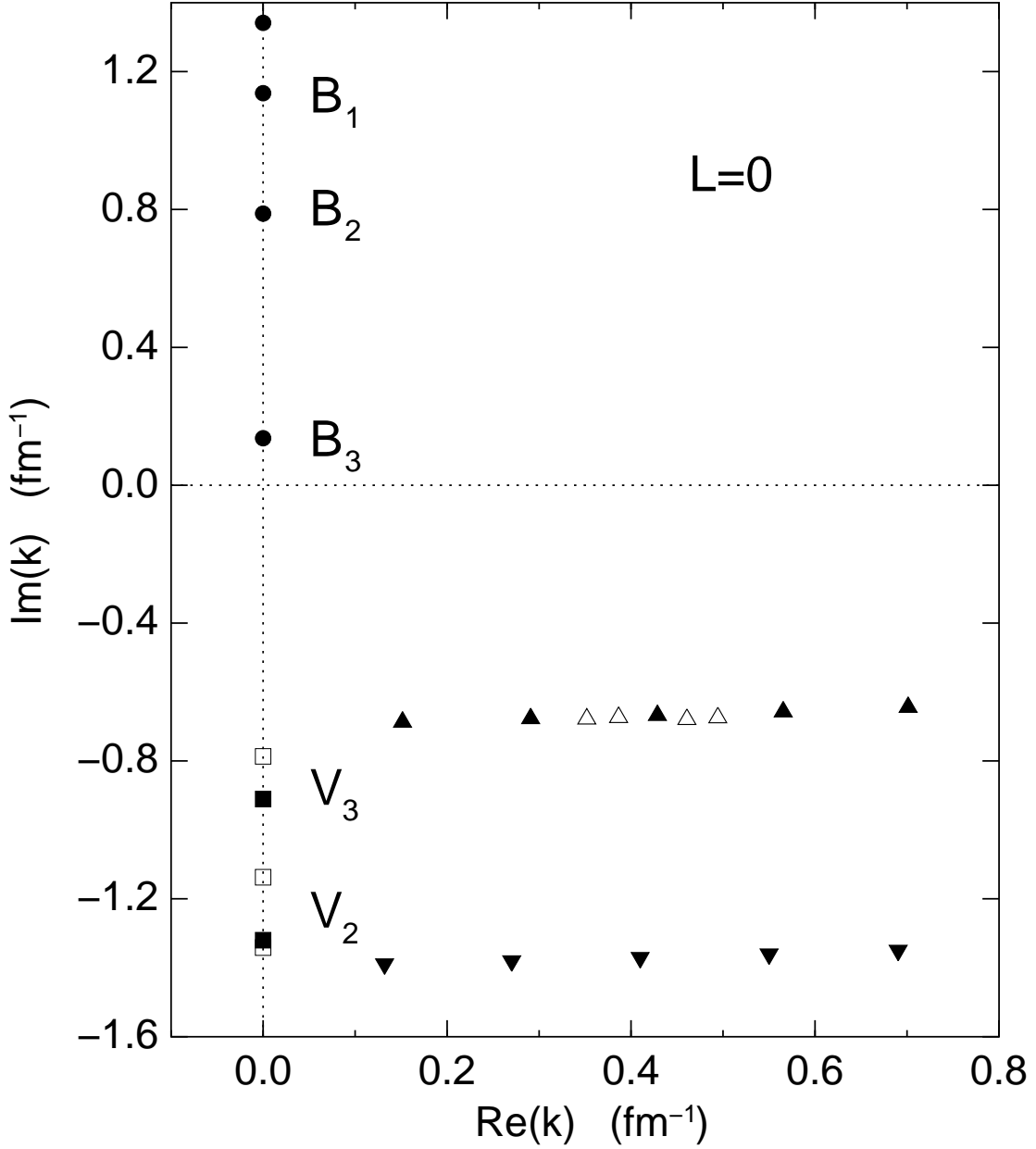}
\caption
{\label{fig2graph}
Eigenvalues for L=0. The bound states are given by full circles, the
virtual bound states by squares, the empty squares are provided by the
free-state boundary condition, while the full squares by the iterated
one. The triangles are the non-physical eigenvalues, the inverse
triangles are provided by the iterated condition. The sensitivity of
the eigenvalue at $k=0.4286-i \cdot 0.6675 \; \mathrm{fm}^{-1}$ on
$R_M$ is demonstrated by the open triangles: they correspond to
$R_M=25, 27.5, 32.5 \; \mathrm{and} \; 35.0 \; \mathrm{fm}$. }
\end{figure}

The eigenvalues received by matching the complete external function to
the regular wavefunction calculated in the internal region exhibit a
suspicious horizontal row just inside the border of the stability
region, cf. fig.\ref{fig2graph}. For the Saxon-Woods potential these
are  not "physical" resonances,
their distance from each other is inversely proportional to
$R_M$. Since in numerical sense the integration is accurate for such
$k$ values, it is only the truncation of the potential, i.e. the
free-state boundary condition imposed at finite $R_M$, which can be
inadequate. This problem shall be studied in detail.

For nonzero L values interesting phenomena can be studied even
with the above simple form of the numerical method. But to make the
logical structure of the present paper more transparent, in the next
section the elimination of the truncation poles is attempted
by improving the
supposed asymptotical behavior. A more detailed exploration of the
lower half of the $k$ plane is left to the next but one section.

\section{ \label{siterb} Long range effects for a short range
potential, iterated  boundary condition }

It was found in the previous section that the free-state boundary
condition imposed at some finite $R_M$ value generates unphysical
eigenvalues, the position of which strongly depends on $R_M$ (cf.
fig.\ref{fig2graph}). In this section a correction term is derived for
L=0,  which
describes the singular behavior at $k=-i \mu/2$ and can be used as a
reference function for the $\mathrm{Im}(k) > -\mu$ region.

In scattering theory it is usual to rewrite the differential equation
for the radial problem  in a Volterra-type integral form \cite{N66}.
The advantage is that it contains the boundary condition too.
The equation with the $\exp(i k r)$ boundary condition at infinity is
\begin{equation}
\begin{split}
\varphi(r) &= \exp(i \cdot k r) + \frac{i}{2 k} \times \\
 & \times \biggl(
\exp(i \cdot k r) \int \limits_r^\infty \exp(-i\cdot k x) V(x)
\varphi(x) \, dx  -
\exp(-i \cdot k r) \int \limits_r^\infty \exp(i\cdot k x) V(x)
\varphi(x) \, dx  \biggr)
\end{split}
\end{equation}
Of course, this equation is meaningful only if
$ \mathrm{Im}(k) > -\mu/2 $.
If only the exponential tail of the Saxon-Woods
potential is taken into account, the first iteration of eq.(5)  is
\begin{equation}
\varphi (r) \rightarrow
\varphi (r)_a = \exp (i \cdot k r)
\biggl( 1+\frac{W_o}{\mu (\mu -i \cdot 2 k)} \exp(-\mu r) \biggr)
\end{equation}
Strictly speaking, this formula is derived with the above
limitation on $k$,
nevertheless one can heuristically consider it in a larger region.

The iterated asymptotics is singular at $k=-i\mu/2$, which corresponds
to the well known singularity of scattering in exponential potentials
\cite{N66}. Since for the eigenvalue problem the normalization is
irrelevant, the pole is considered to be automatically removed by a
factor of $\mu (\mu -i \cdot 2 k)$. But some kind of singular behavior
still remains: it shall be discussed later.

Previously it was established, that the free-state solution can be
used as reference function for the numerical integration at
$\mathrm{Im}(k) > -\mu/2$. The iterated asymptotics is more
accurate, therefore it is meaningful to use it as a new reference
function. The calculation of the new source term is straightforward,
details can be found in \cite{BV95}. A direct comparison of the
minimal solution at the same energy and the calculated new correction
function with the $u(R_M)=0, u'(R_M)=0$ boundary condition shows that
the latter is rising faster during the inward integration now for
$\mathrm{Im}(k) > -\mu$ (cf. fig.\ref{fig1graph}). Consequently, for
the wavefunction the $ R_M \rightarrow \infty $ limit exists in this
region. The better description of the asymptotical behavior results in
a larger stability region for the new scheme.  It should be emphasized
that it was proved only that the new reference function correctly
describes the asymptotics of some solution in a wider region, but
absolutely no information is received whether that is the solution
needed for the definition of resonances.

\begin{figure}
\includegraphics{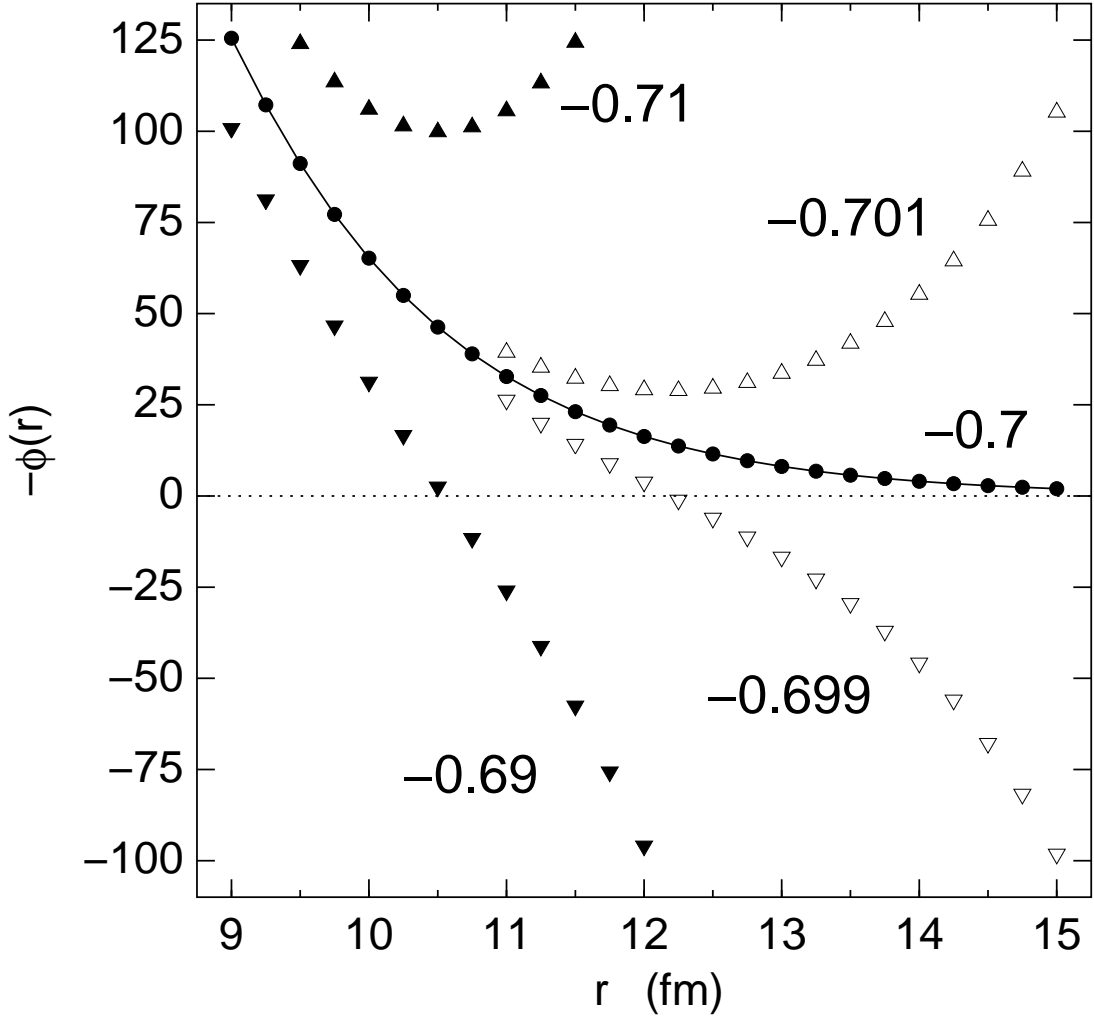}
\caption
{\label{fig3graph}
Behavior of the complete external wavefunction around the range
singularity at $k=-i\mu /2$. The numbers at the curves give the
value of $k$ on the imaginary axis. }
\end{figure}

When the singularity on the imaginary $k$ axis at $k=-i\mu/2$
is passed from above, a zero of the wavefunction moves to infinity and
below it returns as a minimum (cf. fig.\ref{fig3graph}). For every
finite $R_M$ there is an interval when not the rising free-state
solution describes the behavior there: it is a typical long-range
effect. Obviously,
it cannot be described if any approximation is introduced
which alters the the potential around infinity. Note that not
only the structure of the solution is changed, i.e. a node disappears,
but in the singularity the boundary condition coincides with that of
a bound state with the same energy. Some consequences of this
anomalous behavior shall be studied in the next section.
It is interesting that despite the presence of a
singularity, the calculated wavefunction varies smoothly at $R_m$;
even the disappearing zero crosses sufficiently earlier.

The free-state and the iterated boundary conditions give nearly the
same matching condition for
$\mathrm{Im}(k)\gtrsim-0.5\:\mathrm{fm}^{-1}$.
At $k=-i\,\cdot\,0.6\:\mathrm{fm}^{-1}$ the difference between the two
cases is still only 4\% (cf. fig.\ref{fig1graph}).
But in the immediate neighborhood of
$k=-i\,\cdot\,0.7\:\mathrm{fm}^{-1}=-i\mu/2$ the two wavefunctions
behave quite differently. Note that it is impossible to prove in a
strict mathematical sense which case is "correct", since it concerns
the definition of the eigenvalue problem. Only a careful analysis of
the consequences combined with physical intuition can guide us.

To make the comparison of the two approaches easier, the
eigenvalues calculated with the iterated boundary condition are also
given in fig.\ref{fig2graph}. The horizontal row of the unphysical
resonances now
pushed further downwards to the border of the new stability region.
The sensitivity to the value of $R_M$ remains, of course. It means
that one can consider the iterated boundary condition "better" than
the free-state one, but the subtle problem of the proper boundary
condition for resonances is not solved by the iteration.

\section{\label{sexpl} Exploration of the lower half of the $k$
plane.}

In this section some factual results received with the "approximative"
boundary conditions studied so far are collected.
These results are
not directly related to the problem of the correct
asymptotics, nevertheless they are interesting in themselves and they
complement to the general picture. Due to
limitations on space, only some aspects can be discussed, even such
important phenomenon as the existence of the centrifugal eigenvalues
is presented only fragmentary.

\subsection{Node numbers and the system of bound and virtual
bound states.}

The node number is defined only for real valued functions i.e. only
for wavefunctions on the imaginary $k$ axis. It is straightforward to
count the zeroes. For bound states the well-known rule is that the
ground state has no zero, with increasing energy for every next
bound state the node number is increased by one. The $\varphi(\infty)
= 0$ boundary condition for the regular solution can be interpreted
as the appearance of a zero at infinity at that energy, therefore
bound states divide the negative energy axis into sectors, where the
regular solutions have equal node numbers. Because the node at
infinity (and possibly at the origin) is not counted by
definition, the bound state lies at the upper edge of the
corresponding sector. A virtual bound state is also regular, its
node number is defined by the sector corresponding its energy.

When the strength of the potential is increased, bound and virtual
bound states usually arise in a collision of a resonance state with
its time-reversed pair, i.e. with that state which is located
symmetrically to the imaginary $k$ axis. For $L=0$ the collision takes
place on the negative imaginary axis, in other cases in the origin.
Just after the collision the states have equal node numbers. In case
$L=0$ two virtual bound states are formed, one of them moves downward
on the imaginary $k$ axis, while the other moves upward and after
crossing the origin becomes a bound state. Because of the sector
structure, the energy of the corresponding virtual bound state is
always lower than that of the bound state. In some sense the virtual
bound state is locked in its sector: the regularity in itself defines
the solution, normally the energy of a bound state cannot coincide
with the energy of a virtual bound state, consequently it is not
possible to move into the neighboring sector in a simple way. Note
that such a behavior is independent of the the virtual bound state
asymptotics, it entirely follows from the regularity condition in the 
origin.
In the following subsections two cases shall be studied when specific
circumstances make possible for the virtual bound state to leave its
native segment.

\subsection{Exploration of the lower half of the complex $k$ plane for
L=4.}

The modification of the numerical scheme with the free-state reference
function  to nonzero $L$ values is straightforward:  well-known Hankel
functions replace the simple exponential behavior. Despite the
restriction to $\mathrm{Im}(k) \geqq -\mu/2$ wave-number values, some
interesting phenomena can be studied.

\begin{figure}
\includegraphics{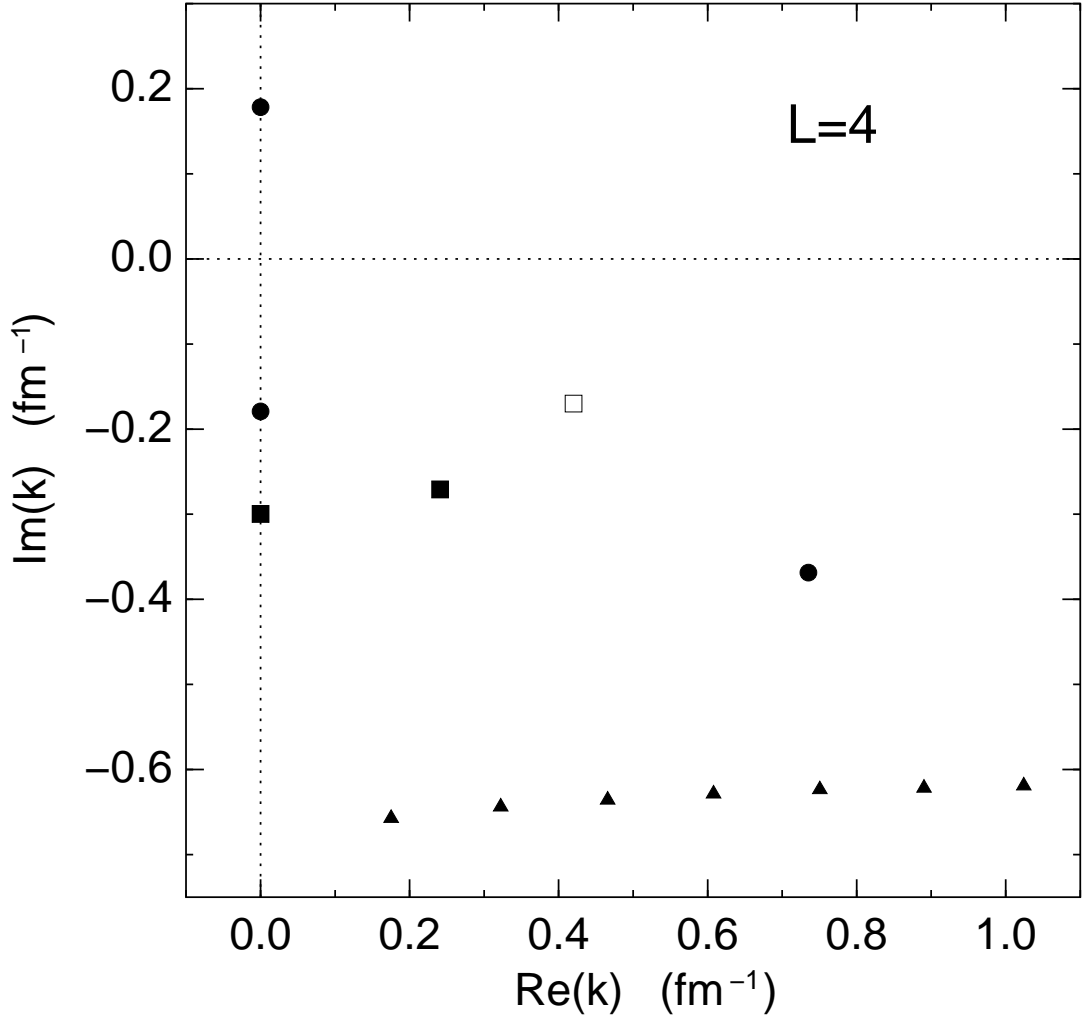}
\caption
{\label{fig5graph}
Eigenvalues for L=4 with free-state asymptotics. "Ordinary" (i.e
bound, virtual bound and resonant in the usual sense) eigenvalues are
given by full circles, the ground state at $i \cdot 0.904 \;
\mathrm{fm}^{-1}$ is not presented. The "unphysical" eigenvalues are
given by triangles, the "centrifugal" ones by squares. At the empty
square an eigenvalue is expected from similar results for the pure
square well potential, but it was not found here. }
\end{figure}

\begin{figure}
\includegraphics{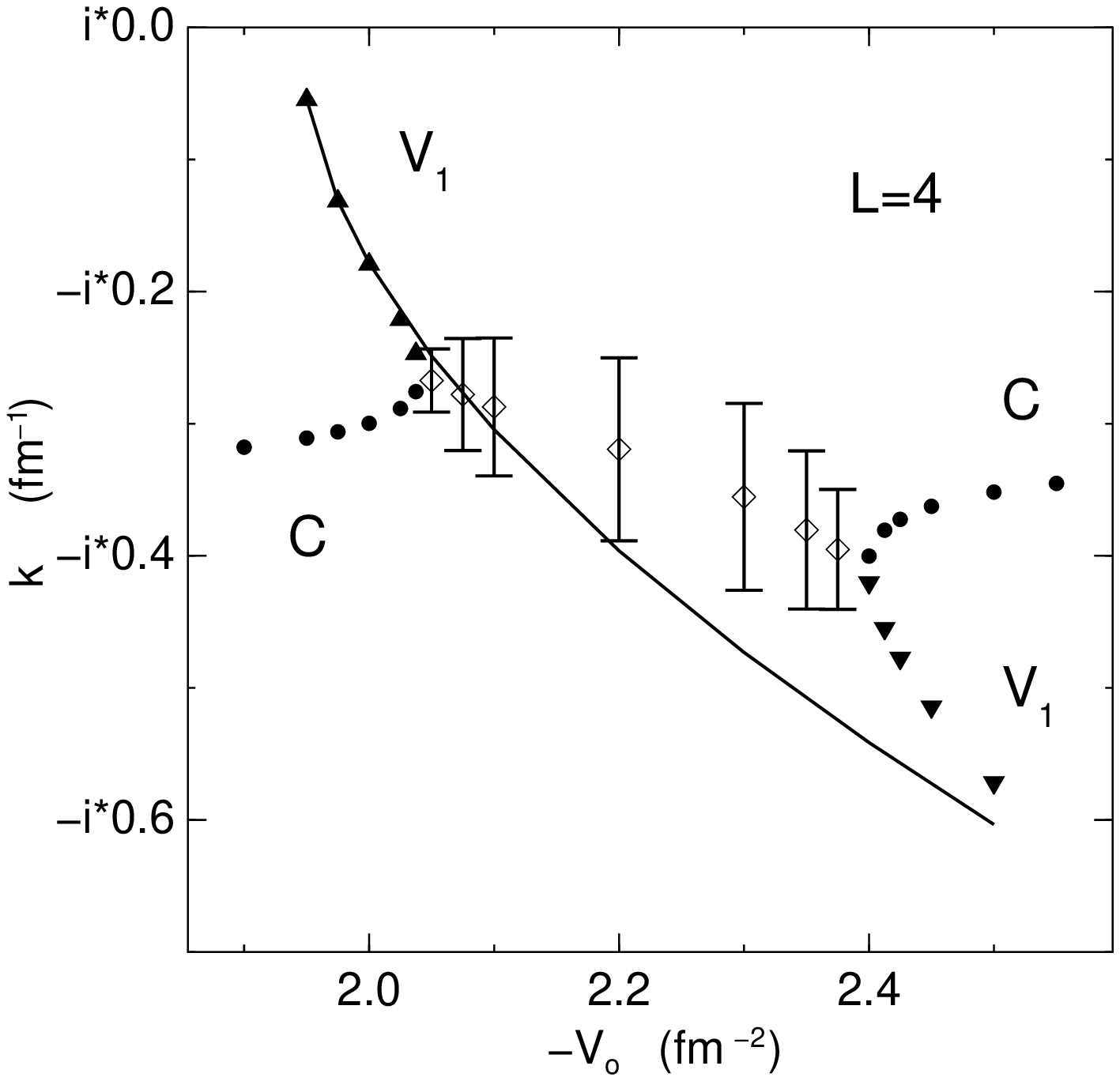}
\caption
{\label{fig6graph}
Eigenvalue trajectories and the centrifugal barrier eigenvalue
on the imaginary axis, L=4.
The virtual bound state pair of the first exited state is given by
triangles, the centrifugal eigenvalue by circles. The solid line
corresponds to the exited state and gives the boarder of the sectors
with one and with two node-numbers. While the colliding eigenvalues
are off
the imaginary axis, the imaginary part of their position is given by
diamonds, an error bar defines the real parts of their positions. }
\end{figure}

The eigenvalues calculated for a relatively large orbital momentum,
$L=4$, are presented in fig.\ref{fig5graph}. By changing different
parameters of the potential one can study their nature and
divide them into different groups.
Besides the bound and virtual bound states, the horizontal row
of equidistant eigenvalues generated by the free-solution boundary
condition appears here too. In addition, the position of some other
eigenvalues depends predominantly on the $R_o$ radius parameter and
is practically insensitive to the $V_o$ strength of the potential.
Obviously, these eigenvalues are generated by the deep potential well
behind the centrifugal barrier, therefore one can call them
centrifugal eigenvalues. And there are eigenvalues, (for instance, in
fig.\ref{fig5graph} the one at $k=0.74-i\,\cdot\,0.37 \;
\mathrm{fm}^{-1}$ ), which
along a power-like trajectory approaches the imaginary axis when the
strength parameter is increased, collides there with its symmetric
pair and one member of the colliding pair becomes the newest bound
state, while the other changes into a virtual bound state. With
increasing strength this virtual state moves downward the negative
imaginary axis, cf. fig.\ref{fig6graph}.

But for \emph{nonzero even}  $L$ values one of the centrifugal
eigenvalues is located on the imaginary axis itself and disturbs the
movement of the virtual bound state. In fact, they collide, leave the
imaginary axis, describe a small loop in the complex plane, after
returning to the imaginary axis they collide again. One of them
remains near the previous place while the other continues its way
downward the axis: the complicated choreography is needed for
changing the node number of the wavefunctions. When the virtual bound
state approaches the centrifugal barrier eigenvalue, the two states
have equal node numbers. After the collision they leave the imaginary
axis, but somewhat later they return. Off the axis the wavefunction is
complex, there are no nodes. While they are off the axis, the bound
state on the positive imaginary axis crosses this energy region,
therefore when they return, they are located in a different sector,
i.e. their node number is increased by one. Now the centrifugal
barrier eigenvalue is "ready to face" the next virtual bound state
with an increased node number. Note that the virtual bound state
continues its way down the imaginary $k$ axis, but now on the other
side of the corresponding bound state, i.e. with a increased node
number. This example illustrates how the eigenvalues are organized
into a system.

\subsection{ Eigenvalue structures for different boundary conditions,
L=0. }

This case provides a possibility to illustrate that the boundary
condition for the wavefunction can be chosen
independently of the reference function, as it has been
already discussed in section \ref{srefm}.

With the free-state solution as the reference function one can stably
integrate the equation only for the $ \mathrm{Im}(k) \geqq -\mu/2$
region. To achieve stability for the larger
$ \mathrm{Im}(k) \geqq -\mu$ region one has to use the
iterated asymptotics.
To find the eigenvalues with the free-state boundary condition
in the larger domain, it is possible to choose the boundary condition
for the correction function appropriately . I chose the
$u(R_M)=0, u'(R_M)=W_o \mu \exp((i k-\mu)R_M)$ condition,
the results for the larger region are also presented in fig 2. The
virtual bound states are locked into their own native segments and
there is a tendency to approximate the lower bound state. In
fig.\ref{fig2graph} all of the three free-state boundary condition
defined virtual bound states given by empty squares are extremely near
to these bound states.

\begin{figure}
\includegraphics{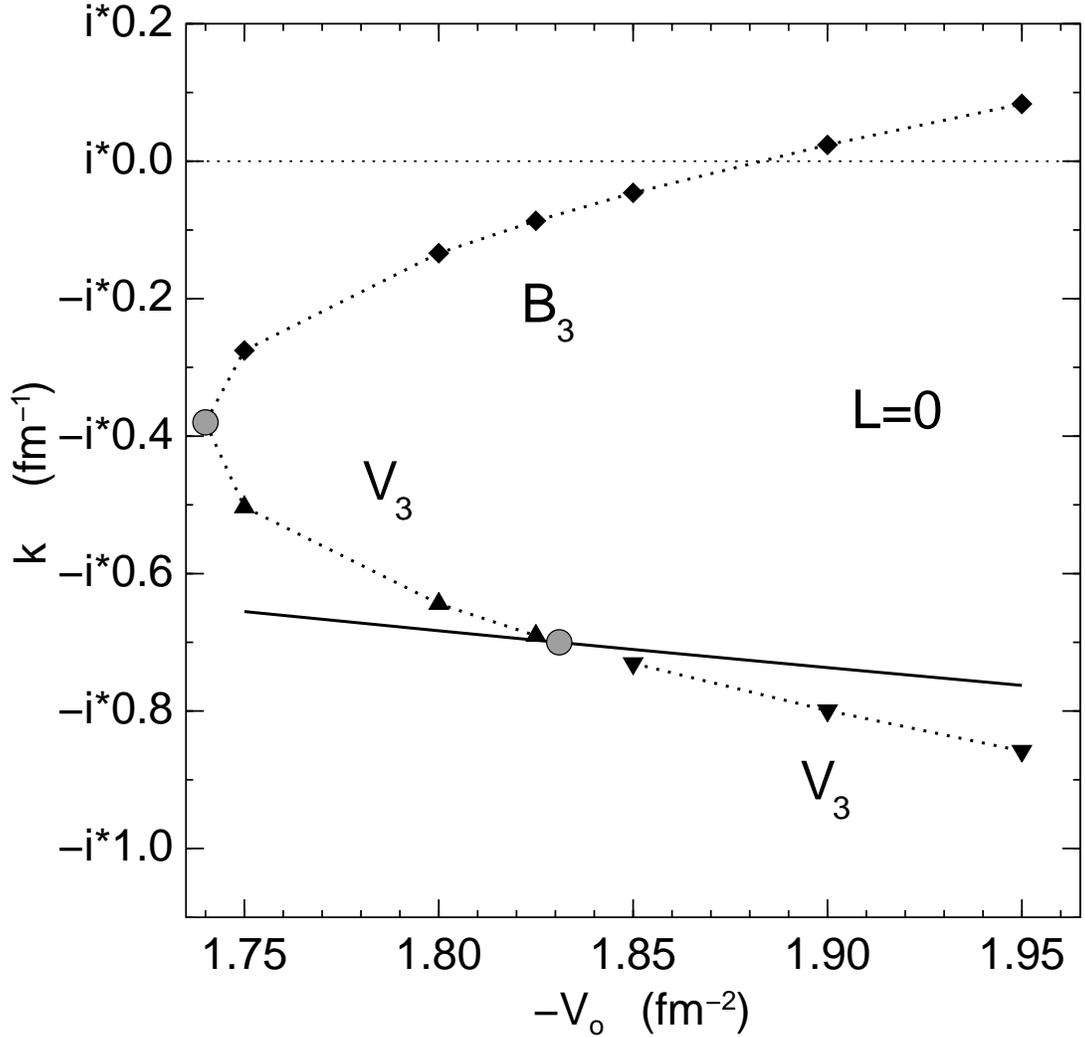}
\caption
{\label{fig7graph}
Eigenvalue trajectories and the iterated singularity, L=0. The state
which becomes the third exited state is given by diamonds, its virtual
bound state pair by triangles. The solid line gives the segment
boarder defined by the second exited state. Large half filled circle
denotes the area where entering of $\mathrm{V}_3$ into the segment
of $\mathrm{B}_2$ takes place; during the crossing even the
wavefunctions coincide. A similar circle denotes the formation of
$\mathrm{B}_3$ and $\mathrm{V}_3$ in a collision of two resonances. }
\end{figure}

In case of the iterated asymptotical behavior, the boundary condition
for the virtual bound state in the singularity at $k=-i\mu/2$
is practically the
same as for the bound state, therefore in this case the two states can
coincide. Moreover, they should coincide: a bound state eigenvalue
there implies a virtual bound state too, and vice versa. It is
illustrated in fig.\ref{fig7graph} that a bound state reaches the
energy of the singularity simultaneously with a virtual bound state,
the virtual state sheds a zero, leaves its native segment and enters
the segment defined by the bound state. This is an other example of
how a virtual bound state can change its node number.

Therefore the virtual bound states can be extremely
sensitive to the asymptotical behavior of the potential. In
fig.\ref{fig2graph} the states which correspond to the fourth bound
state $\mathrm{B}_3$  and which are
located at $k=-i \cdot 0.79 \, \mathrm{fm}^{-1}$ and at $k=-i \cdot
0.91 \, \mathrm{fm}^{-1}$ seem to be not far from each other,
nevertheless they are completely different: the node numbers are 3 and
2.

The ground state arises not in a collision of two resonances,
but the corresponding eigenvalue
is generated directly as a virtual bound state while the
potential is still extremely weak. For example, at $V_o=-0.0001 \,
\mathrm{fm}^{-2}$ there is a virtual bound state at $k=-i \cdot 0.51
\, \mathrm{fm}^{-1}$. At the same time, there is no other eigenvalue
on the imaginary axis in the stability region of the numerical method.
To avoid any misunderstanding, it should be emphasized that such a
weak potential generates only a small perturbation of the internal and
external wavefunctions, typically it alters only the fifth digits. The
eigenvalue arises as the concurrence of these slightly perturbed
wavefunctions. With increasing strength of the potential this virtual
bound state moves towards the origin and eventually becomes the ground
state.

In contrast, the presence of the centrifugal barrier for $L \neq 0$
modifies the genesis of the ground state. At very week potential only
the more or less fix centrifugal eigenvalues arise, the ground state
comes from a pair which collide at $k=0$.

\section{\label{sanal} Analytical behavior of the solutions}

It is timely to look in perspective at how the asymptotical behavior
is chosen for the eigenvalue problem. Because of the stability, for
bound states different possibilities are feasible. It is usual to
suppose that the exponentially decreasing free-state solution
provides the asymptotics. But note that even the $\varphi (R_M) = 0$
boundary condition can be used.
The minimal solutions also provide a more sophisticated
asymptotics.

On the lower half-plane resonances and virtual bound states are,
as a rule, instable, the free-state boundary condition is not
satisfactory. An improved approximation of the asymptotical behavior 
was introduced, it proved to be inadequate too: the unphysical
eigenvalues are not eliminated. Therefore the iteration process can
not solve the principal problem of the proper boundary condition.
It seems obvious that the minimal solutions contain in some form the
information still missing from the iterated asymptotics, but
unfortunately these are defined only on a different half-plane.

\subsection{Principle of analytical dependence of the solutions}

At different energies the minimal solutions are completely
independent: their normalization is arbitrary. Therefore the values at
fixed $r$ do not provide an analytical function of the $k$ variable.
Even if such a normalization is used which provides analytical
values, in the generic case the derivatives are not necessarily
analytical. But if both the values and the derivatives happen to be
analytical, this behavior is preserved when the solutions are
integrated to a different $r$ because the Schr\"odinger equation
contains the $k$ parameter in an analytical form.  Unfortunately, I
cannot give due reference in the literature to this "obvious"
statement.  In some sense minimal solutions can be considered to
be the limits of solutions with the $\varphi(R_M)=0, \varphi'(R_M) =
\text{const.}$ boundary condition. Therefore, if the normalization
constants are correctly chosen, minimal solutions provide an
analytical function of the $k$ parameter. Of course, the choice is not
unique, but it concerns only the normalization, a property irrelevant
for the eigenvalue problem. Once an analytical dependence is achieved,
the solutions can be continued to such regions of the $k$ variable
where they were originally not defined. This continuation is
essentially (i.e. apart from the normalization) unique.

On the analytical dependence a constraint can be imposed. For
real potentials and for real energies the solutions can be chosen to
be real. Usually, the minimal solutions at negative energies are
considered to be real, I follow this practice. Because
they are real, at some fixed $r$ value the wavefunctions satisfy the
basic constraint

\begin{equation}
\varphi(k,r)=\varphi^{\ast}(-k^{\ast},r)
\end{equation}
where  asterisk denotes the complex conjugation. Since not only the
left hand, but also the right hand side is an analytical function of
the $k$ variable, the equation holds everywhere. In presence of branch
points, however, one should carefully follow the possibly different
sheets of $k$ and $k^{\ast}$.

The just formulated basic constraint is in fact the
consequence of time reversal symmetry. But when Wigner studied
time reversal (cf. ref.\cite{Wig}, for instance) only real physical
quantities were considered. Therefore, he derived the $\varphi
\rightarrow \varphi^{\ast}$ and the $ k \rightarrow -k$ transformation
rules. The later differs from the $k \rightarrow -k^{\ast}$ rule
suggested by the above formula. The extra complex
conjugation is the principal reason that time reversal does not
help in defining a solution different from the minimal one, since it
maps the minimal subspace onto a minimal subspace at different energy
rather than into the solution space at the same energy.

\subsection{Continuation of the minimal solutions}

 The free-state solutions describe the asymptotical
behavior not only for the minimal solutions, but they provide
sufficiently accurate asymptotics on the real $k$ axis too.
Therefore they were also used for the
boundary condition in the lower half-plane. Note that the
evaluation of an explicit formula with an argument in the lower
half-plane is equivalent with an analytical continuation. It means
that a continuation procedure has been
already applied. But  somewhere the continuation of an
approximation can considerably deviate from the continuation of the
function itself. A better approximation,
the iterated asymptotics, gave satisfactory results in a larger
domain. It needs, however, calculations using information on the
potential itself and resulting in explicit formulas to be able to
perform the continuation.

A convenient way to deal with this difficulty can be just the
opposite approach: a continuation based on "raw" numerical values of
the minimal solutions accurately calculated, let us say, at $R_m$
rather than at $R_M$. It saves even the inward integration and
provides directly the matching condition.

In approximation theory of analytical functions \cite{W60} the
phenomenon of "superconvergence" is well known. If a function is 
considered on a domain, the rate of convergence for any
possible polynomial approximation sequence is limited by the nearest
singularity. A sequence reaching the possible fastest rate is called
converging "maximally". The base theorem is that any maximally
converging polynomial sequence converges to the continued function in
a larger domain, of course, with a smaller rate of convergence.  By
the geometry of the domain and by the nearest singularity this rate is
defined in a relatively complicated manner (cf. chapter IV in
ref.\cite{W60}). Note that the superconvergence theorem provides
no effective error estimate, the accuracy should be checked by other
means.

Exploiting the phenomenon of superconvergence  an "empirical"
continuation was applied to measured nuclear reaction data (cross
sections and polarizations) to extract structure information in a
model independent way \cite{B76,BG89}. The interested reader is
referred to the cited literature, \cite{B76} is perhaps the first
choice. But to understand the present paper all that subtlety of
convergence rate estimates and optimal conformal mappings together
with the various methods to check the reliability of the results are
not needed. 

To illustrate the possibilities, numerical results
provided by the iterated boundary condition are used with
$R_M=35\,\mathrm{fm}$ and $V=2.0\,\mathrm{fm}^{-2}$.
The solutions at $R_m=7.5\,\mathrm{fm}$ are defined by two ordered
complex numbers, $\varphi(k,R_m)$ and $\varphi'(k,R_m)$, one of them
is nonzero and the normalization is irrelevant. One's first guess is
to fit $\varphi/\varphi'$ with a rational fraction $\mathrm{P}_n(k) /
\mathrm{Q}_m(k)$, where $\mathrm{P}_n(k)$ and $\mathrm{Q}_m(k)$ are
polynomials of order $n$ and $m$. The difficulty with this approach is
that the result is not symmetric, i.e. a fit to $\varphi'/\varphi$
with $\mathrm{Q}_m(k) / \mathrm{P}_n(k)$ gives a different result. The
reason is that in the complex plane the distance of $\varphi/\varphi'$
and $\mathrm{P}_n(k) / \mathrm{Q}_m(k)$ is different from the distance
of $\varphi'/\varphi$ and $\mathrm{Q}_m(k) / \mathrm{P}_n(k)$. A
solution to this problem is presented in the appendix. Maximal
convergence can be achieved using somewhat different approximation
criteria, therefore other possibilities are also feasible.

\begin{figure}
\includegraphics{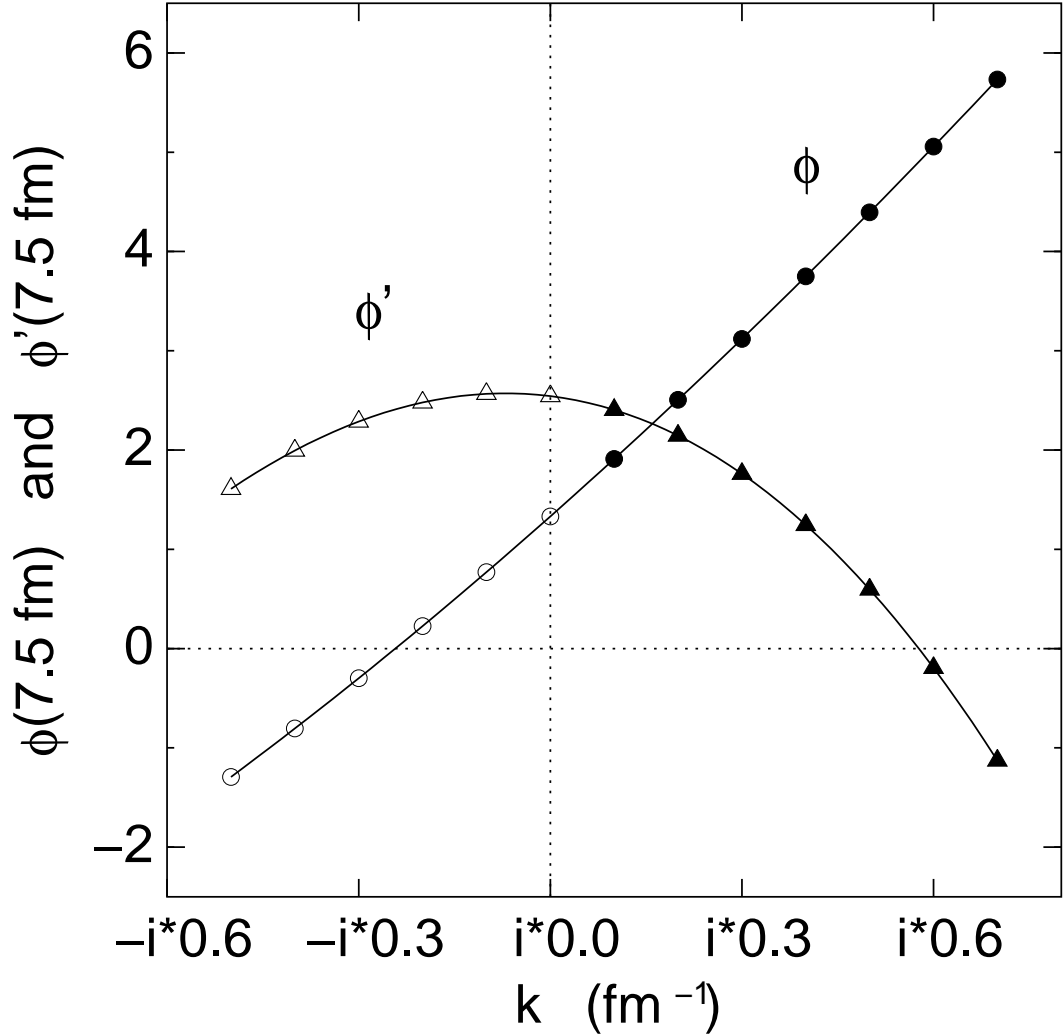}
\caption
{\label{fig4graph} Extrapolation of the minimal solutions to the lower
half-plane. The calculated values for the complete wavefunction $\phi$
and its derivative $\phi'$ at $R_m$ on the imaginary $k$ axis are
given by circles and triangles, respectively. The solid lines are
fitted fifth degree polynomials; the full symbols give
fitted values, they were fitted together with other ones in
the complex $k$ plane; the empty symbols give values calculated for
checking the quality of the extrapolation. For details see
the main text.}
\end{figure}

In the $k$ plane an equidistant grid was used at
$k=0.1\cdot( n+i\cdot k )$ with $n,k=0,1,2,\dotsc$. The fitted points
were chosen on the border and inside a triangle defined by three
points $ 0.0 + i\cdot 0.1$, $ 0.0 + i\cdot 0.6$ and $ 0.6 +
i\cdot 0.1$, altogether 28 ones. The used large $R_M$ value assured
that in these points the minimal solutions themselves were calculated.
The fitting polynomials are of the fifth degree, it needs 11 fitted
real parameters (because the functions are real for imaginary $k$).
The relative accuracy of the fit is typically $10^{-5}$ inside the
triangle, $5\cdot10^{-5}$ in its peaks. The performance of the
continuation to the lower half-plane was checked on the imaginary axis
by comparing to calculated values (cf. fig.\ref{fig4graph}). Even in
the farthest
point the accuracy is $10^{-3}$! It might be surprising, nevertheless
nothing unexpected happened.

First of all, the base region for the extrapolation (i.e. the region
where the fit was performed) contains 28 accurately calculated
points, its dimensions just equal the distance of extrapolation. It
means that large amount of information was used and the distance of
extrapolation is moderate. Secondly, the circumstances are favorable.
Apparently, there is no singularity nearby, in the singularity at
$k=-i\mu/2$ the function which was extrapolated is regular 
and the next possible singularity is at $k=-i\mu$. On the whole,
one can expect such a good performance, nevertheless the small number
of parameters needed is quite promising.

The performed check provides
information on the iterated boundary condition too. The facts that 
practically the minimal solutions themselves are continued 
and the good agreement
with the calculated values in the lower half-plane indicate that in
this region one can apply the iterated boundary condition. With other
words, a mutual check is performed.

I do not want to make the impression that continuation is a
simple procedure. The idea that one should analytically continue some
function into an other region for solving different problems in
physics is not new. A quite early example can be found in
ref.\cite{Ch58}, for instance. There has been numerous attempts since
then, most of them failed in some sense or another.  There are some
pitfalls, irresponsible applications not taking them into account and
not checking the results carefully can do much harm (c.f. some
applications reviewed in refs.\cite{B76,BG89}).  Due to limitations on
space, no introduction to this complicated subject can be given here. 
Basically, there are four possibilities for checking the performance. 
First of all, the shape and position of
the base region for the continuation can be altered, the independent
variable can be changed using a suitably chosen conformal mapping, the
contribution of the nearby singularities can be suppressed with
multiplicative factors, and finally one can examine the performance of
the continuation in such directions where direct calculation is still
possible. 
In this respect the fact that the function to be continued can be
accurately calculated practically in a whole half-plane provides
unique possibilities. But there is no
need to apply any of them here, since a direct check was performed.

\section{\label{novel} Implications for scattering calculations: a
novel approach}

For short range potentials the so called "outgoing" solutions are the
continuation of the minimal solutions to the positive part of the real
$k$ axis, while the "ingoing" solutions are the continuation to the
negative part. For long range potentials the continued functions can
be considered as the definition of the ingoing and outgoing solutions,
for pure Coulomb potential the known solutions satisfy this
condition. It means that the calculation of scattering processes
was implicitly also treated above. For it the regular wavefunction of the
internal region at $R_m$ should be described as a linear combination
of the outgoing and ingoing solutions rather then only one of them, as
in the case of the eigenvalue problem. The coefficients of the linear
combination determine the S-matrix, to be exact, the so called
"distorted" S-matrix. The basic constraint imposed on the minimal
solutions assures that the normalization ambiguity has no effect,
the Wronskian for the in- and outgoing solutions can also be used to
fix the normalization. In this way the S-matrix can be easily computed
even for quite "exotic" interactions, the only prerequisite condition is the
existence of minimal solutions.

The connection between the minimal and the in- and outgoing solutions
shows that an eigenvalue defined with the continued minimal solution
boundary condition coincides with a pole of the S-matrix.  The
S-matrix is a ratio of two coefficients, a pole arises when a zero of
the denominator occurs. It is just the definition of the eigenvalue:
in the asymptotical behavior only one of the linearly
independent minimal and continued minimal solutions is present.

For short range potentials with their free-state asymptotics it is
straightforward to calculate "observable quantities" from the
S-matrix. But long range potentials influence the kinematics of the
scattering, to be able to calculate observables one has to know the
"physical interpretation". Such an interpretation
is performed when the complete wavefunction is separated into the
"incoming" and the "scattered" components, while the physical
quantities are calculated from them.
A clear illustration of it is provided by the standard treatment of
the Coulomb case. As a result, observables are defined not only by
the "Coulomb distorted S-matrix", but also by "Coulomb phases".

Hopefully, these considerations help to take into account Coulomb
interactions quite accurately also in three-particle calculations. The
leading terms (and only the \emph{leading} terms) of the asymptotics
for various components are well studied, a somewhat detailed
discussion can be found in ref.\cite{VKR01}. The characteristic
exponentially rising and decreasing behavior is present, therefore
minimal-type solutions should exist. But now the solution space is not
two-dimensional, i.e. its structure is more complicated, a stable
inward integration of a minimal-type solution is not so simple and
straightforward. Therefore, much has to be done to implement
the new scheme in a three-particle calculation, only the main steps
are clear.

\section{\label{sconcl} Conclusions}

In the present paper resonances and virtual bound states are studied 
by considering the eigenvalue problem for the Schr\"odinger
equation. It requires the correct boundary
condition for the formulation and an accurate numerical method to
solve the equation.

A new and effective approach, the reference method, is developed 
to solve the equation. Even in its simplest form, i.e. with the
free-state reference function, it is suitable to explore the lower
half of the complex $k$ plane, in the neighborhood of the imaginary
axis too. Keeping in mind the extreme simplicity of the numerical
approach, one wonders what is the reason that it was proposed only
quite lately and why it is not the standard method for solving the
eigenvalue problem, let us say, from the mid-sixties of the last
century. Note that even the complex rotation method, usually
considered to be very powerful, is not suitable to calculate virtual
bound states.

The reliable results provided by the method reveal that not the
free-state asymptotical behavior defines resonances and virtual
bound states, even for short range potentials. The free-state behavior
should be considered only approximative, it is more or less
accurate only near the real $k$ axis. The existence of eigenvalues
depending on $R_M$ clearly shows it. The first iteration of the
Volterra-type integral form of the Schr\"odinger equation provides a
correction to the free-state boundary condition. Numerical
calculations show that the unphysical eigenvalues survive, only their
position is modified.

Despite the approximative nature of the boundary conditions
discussed so far, the reference method can provide interesting
information. The lower half of the $k$ plane was  explored, some
results are collected in a separate section. Since the bound and
virtual bound states satisfy the same boundary condition at the
origin, their position is not independent. Therefore the system they
form was studied and also how it changes with increasing strength of the 
potential.

Nearly fixed eigenvalues describing the potential well behind the
centrifugal barrier were found. For nonzero even orbital momenta
a complicated choreography of collisions, in which a virtual bound 
state and the centrifugal eigenvalue take part, was observed .
In the collision process the node numbers are increasing.

On the contrary, the virtual bound state sheds a node when crosses the
range singularity generated by the exponential tail of the potential.
Moreover, in the moment of crossing its wavefunction  coincides with
the wavefunction of a bound state at the same energy. Such an exotic
behavior can influence the nuclear surface.

To find the correct boundary condition, it is necessary to study the
structure of the solution space. It is straightforward to define in a
simple and natural way the minimal solutions. They provide the
natural boundary condition for the bound states. But to find an
"other" solution, which can be used for resonances and virtual bound
states, is not straightforward. Neither time reversal nor the
symplectic structure provided by the Wronski determinant can
help. Carefully analyzing the situation one concludes that the correct
condition is given by the analytical continuation of the minimal
solutions from the upper half-plane to the lower one. In some sense
this conclusion is the main result of the present paper: 
one can formulate the
eigenproblem simultaneously for the bound, the resonant and the
virtual bound states.

The principle of analytical dependence provides practical means too.
Numerical information on the minimal solutions was used and 
they were continued 
to the lower half of the $k$ plane, where the results were compared
to the solutions calculated with the reference method using
the iterated boundary condition. In this way it was demonstrated that one
can effectively perform the continuation. Obviously,
the successful continuation of the minimal solutions, i.e. the
matching condition itself, means that one can successfully calculate
resonances and virtual bound states too.

The importance of the continuation procedure lies not in
making obsolete the integration with the aid of the reference method.
In many cases the free-state reference function (or its generalization
in presence of a Coulomb potential) is accurate enough. For
the nuclear structure calculation practice its is usually irrelevant
whether the exponential tail for a phenomenological potential is
correctly taken into account. If phenomena directly connected
to the region just outside of the surface layer (i.e to the 10-13 fm
region in fig.\ref{fig3graph}) are studied, then one can apply the
iterated asymptotics as the reference function. In this way it is
quite safe to apply the simple and transparent reference method  for
the numerical integration, rather then to perform the relatively
unknown continuation. But the latter, i.e. the continuation, is very
useful if no free-state type solutions are explicitly known. At
complex energies one can easily calculate the minimal solutions, their
continuation to the lower half-plane provides the necessary matching
condition.

The analytical continuation approach provides unique possibilities for
scattering calculations. An important new scheme follows from it in a
simple way. Since the in- and outgoing solutions are the continuation
of the minimal solutions to the real $k$ axis, one can represent the
physical solution, which is regular at the origin, as a linear
combination of them, even if the analytical form is unknown. The
"distorted" S-matrix is provided by the ratio of the corresponding
coefficients. This scheme is not restricted to the radial equation,
but to implement it in more complicated cases some technical problems
have to be solved.

Finally, it simply follows from the definitions that the eigenvalues
defined with the analytically continued minimal solution boundary
condition coincide with the poles of the S-matrix.

\appendix

\section{Least squares for the projective complex plane}

As it was discussed above, the solutions are defined by two ordered
complex numbers, $\varphi(k,R_m)$ and $\varphi'(k,R_m)$, one of them
is nonzero and the normalization is irrelevant. This is a typical
example of what is called complex projective plane and denoted by
$\mathrm{P}\mathbb{C}$. If the reader is acquainted with modern
differential geometry, he can easily realize that by the stereographic
projection a diffeomorfizm onto a two dimensional sphere $S^2$ is
constructed and the metric of the embedding three dimensional space
$\mathbb{R}^3$ is used. Below only the algorithm is described.

The elements of $\mathrm{P}\mathbb{C}$ are denoted by (v,w), while
elements in $\mathbb{R}^3$ are given by their coordinates (x,y,z). A
two dimensional sphere $S^2$ defined by $x^2+y^2+(z-1/2)^2=1/4$ is
also considered.

\begin{itemize}
\item If $w \neq 0$ calculate
$( x_1=\mathrm{Re}(v/w), y_1=\mathrm{Im}(v/w), z_1=0)$ and consider
the straight line connecting this point with the (0,0,1) point on
$S^2$. Calculate $( x_v=x_1 t_1, y_v=y_1 t_1, z_v=1-t_1  )$ with
$t_1=(x_1^2+y_1^2+1)^{-1}$. This point is the intersection with the
surface of $S^2$ and gives the result of the mapping.
\item If $v \neq 0$ calculate
$( x_2=\mathrm{Re}(w/v), y_2=-\mathrm{Im}(w/v), z_2=1 )$ and consider
the straight line connecting this point with the (0,0,0) point on
$S^2$. Calculate $( x_w=x_2 t_2, y_w=y_2 t_2, z_w=t_2  )$ with
$t_2=(x_2^2+y_2^2+1)^{-1}$. This point is the intersection with the
surface of $S^2$ and gives the result of the mapping.
\item If neither $v$ nor $w$ equals zero, than because of the
different sign in formulas for $y_1$ and for $y_2$ the two mappings
define the same image point, i.e. it is irrelevant which one is
chosen. Numerical accuracy prefers to divide by that component the
absolute value of which is larger.
\end{itemize}

The last step is to calculate the distance between a fixed and a
fitting point in $\mathrm{P}\mathbb{C}$ as the $\mathbb{R}^3$ distance
of their images on $S^2$. It is a smoother function then the distance
on the surface of $S^2$. As usual, the function to be minimized is
the sum of the squared distances.

\begin{acknowledgments}
Thanks are due to A.\ T.\ Kruppa of Debrecen for providing information
on the complex scaling method.
\end{acknowledgments}

\end{document}